\documentstyle[11pt]{article}

\title{\vspace{4cm}\large\bf
Interplay Between Gravity and Quintessence:\\ A Set of New GR Solutions
}
\author{Arthur D.~Chernin\\
 Sternberg Astronomical Institute, Moscow University, Moscow, 119899,
Russia,\\
Astronomy Division, Oulu University, Oulu, 90401, Finland,\\
and Tuorla Observatory, University of Turku, Piikki\"o, 21 500, Finland\\
David I.~Santiago\\
Department of Physics, Stanford University, Stanford, CA 940305, USA\\
Alexander S. Silbergleit\\
Gravity Probe B, W.W.Hansen Experimental Physics Laboratory,\\
Stanford University, Stanford, CA 940305-4085, USA
}

\date{~}

\begin{document}

\maketitle

\begin{abstract}
\noindent
A set of new exact analytical General Relativity (GR) solutions with
time-dependent and spatially inhomogeneous quintessence
demonstrate 1) a static non-empty space-time with a horizon-type
singular surface; 2) time-dependent spatially homogeneous `spheres' which
are completely different in geometry from the Friedmann isotropic models;
3) infinitely strong anti-gravity
at a `true' singularity where the density is infinitely large.
It is also found that 4) the GR solutions allow for an extreme `density-free'
form of energy that can generate regular space-time geometries.


\end{abstract}
%
%

{\em Introduction. ---}

The idea of quintessence (Q) as a dynamic, time-dependent and spatially
inhomogeneous energy with negative pressure-to-density ratio
($p = w \rho, -1 \le w < 0$) [1] provides new `degrees of freedom' in
cosmology [1, 2] and extends the variety of modern field models
to include extreme forms of energy [1, 3]. It may also stimulate
a better understanding of the fundamental problem of
interplay between gravity and the known field theories
(for a review of this problem, see [4]).

The equation of state with $w = -1$ represents vacuum-type
quintessence (VQ) which
is phenomenologically described by the cosmological constant [5];
VQ is known to induce the
dynamic effect of cosmological acceleration, if its density is positive.
Recent data on the cosmological distribution of distant SN Ia [6] (as well
as the evidence comming from the cosmic age, large scale structure, and
cosmic microwave
background anisotropy combined with the cluster dynamics)
indicate that the observed cosmological expansion is indeed
accelerating,  most probably.
The physical reason for the cosmic acceleration
may generally be attributed not onlty to VQ, but to any form of
Q with $w < - 1/3, \rho > 0$ or $w > - 1/3, \rho < 0$, as seen
from the Friedmann equations. In the original paper [1],
positive energy for Q is preferred; in contrast with that, the whole range
$- \infty < \rho < + \infty$ is considered below, for
completeness and to examine whether any restrictions on the sign
of energy could come from GR, if $w = const$. Note that the
`energy dominance condition', $\rho + p > 0$, which is not met in
the basic case of VQ, $w= - 1$, is not satisfied also for some other
forms of energy discussed below.

The special case $w = - 1/3$ represents the only form of energy (with the linear equation of state) that can generate gravity with zero acceleration
or deceleration effect in an isotropic universe, regardless the sign of the
energy density.  Gravity of
this origin reveals itself only in the curvature of the four-dimensional
space-time, it has no Newtonian analogs (unlike all other forms of
Q, including VQ), and
its nature is completely due to General Relativity (GR) physics.
Historically,
Q with $w = - 1/3$ was the first energy component that appeared in
GR cosmology: in the Einstein static cosmological solution of 1917 the
dynamic balance of anti-gravity of the cosmological constant and
gravity of pressure-free matter gives rise to an effective equation of
state with $w = - 1/3$.
This special type of Q with $w = - 1/3$ may be called {\em Einstein
quintessence} (EQ).

In the present Letter, some special properties of gravity produced by EQ,
as well as by other extreme forms of Q in static and time-dependent,
spatially homogeneous and inhomogeneous space-times are studied
by means of a set of new exact analytical GR solutions. The metric of the
solutions has 3D spherical symmetry,
\begin{equation}
ds^2 = A(r,t) dt^2 - B(r,t) d\Omega ^2 - C(r,t) dr^2;
\end{equation}
\noindent
the componets of the metric tensor $A(r,t), B(r,t), C(r,t)$ are
functions of the
radial coordinate $r$ and time $t$, and $d\Omega^2 = \sin ^2 \theta
d \varphi ^2 + d \theta ^2$.

Four major new results are reported in the Letter.
1) The interplay between gravity and Q is able of creating a
static non-empty space-time with a horizon-type singular surface;
2) This interplay is also revealed in the formation of time-dependent
spatially homogeneous `spheres' which are
completely different from the Friedmann isotropic models and have no analogs
in Newtonian gravity;
3) Q can induce infinitely strong anti-gravity
at a `true' singularity where the density is infinitely large;
4) GR allows for an extreme `density-free' form of energy that can
generate regular space-time geometries. We use the units with $G = c = 1$.

{\em Solution S$\pm$. }

All `static' [no dependence on $t$ in the metric coefficients (1)] solutions to Einstein's equations (see, for instance, [7]) with EQ equation
of state ($w = - 1/3$) are given by
\begin{equation}
B(r)=r^2,\qquad C=\frac{r^4}{a^2}\,\frac{\left(A^{\prime}\right)^2}{A},\qquad
8\pi\rho=\frac{3a^2}{r^5}\,\frac{A}{\left(A^{\prime}\right)^2}\left(\frac{A{
''}}{A^{'}}+\frac{2}{r}\right)\; ,
\end{equation}
\noindent
where $A(r)$ is determined from a quadratic equation
\begin{equation}
kA^2+\frac{2}{a^2}\,A+b+\frac{1}{r^2}=0\; ,
\end{equation}
\noindent
with three arbitrary constants $k$, $b$ and $a^2>0$.
For $k\not=0$ the two (S$\pm$) solutions thus are
\begin{equation}
A(r) =\frac{1}{ka^2} \left[-1\pm\sqrt{K(r)}\right],\qquad
C(r) = \frac{-1\mp\sqrt{K(r)}}{(br^2+1)K(r)}\; ;
\end{equation}
\begin{equation}
8\pi\rho(r)=-\frac{3(1-ka^4b)}{ka^4}\,\left[-1\pm\sqrt{K(r)}\right],\qquad
K(r)=1-ka^4(b+1/r^2)\; .
\end{equation}

To describe the parameter cases, it is convenient to introduce $q\equiv1-ka^4b$, so that
$K(r)=q-ka^4/r^2$. Evidently, $\sqrt{K(r)}$ should be positive at least for some interval
of $r$, which leads to the following:

\noindent A) For $q\leq0$, $k>0$ there are no (physical) solutions 

\noindent B) For $q<0$, $k<0$ we have "interior" solutions, $r^2\leq ka^4/q$;
G+ does not have a horizon for $q<-1/2$, its horizon coincides with the
boundary  $r^2\geq 2|k|a^4$ for $q=-1/2$, there is a horizon inside the
spacetime for $-1/2<q<0$; apparently, S- always has no horizon.

\noindent C) For $q=0, \rho = 0$, $k<0$ S+ is the Schwarzshild
solution, S- is not physical because it is also Schwarzshild but with the
negative central mass.

\noindent D) For $q>0$, $k<0$ both solutions are "global" (spacetime extends to any $r$), S+ has a horizon.

\noindent E) For $q>0$, $k>0$ both are "exterior" solutions, $r^2\geq ka^4/q$,
S+ has no horizon for $0<q<1/2$, its horizon coincides with the boundary
$r^2\geq 2ka^4$ for $q=1/2$, there is a horizon inside the spacetime for
$q>1/2$.

Note that whenever $r=0$ is within the spacetime, solutions S$\pm$ have there a true (non-coordinate) singularity: both the Ricci curvature and the density are infinite at $r=0$. For S- this is a naked singularity, with no horizon around it. Note also that in the lomit $k\to0$ S+ goes to the solution S-1, which is obtained from Eq.(3) with $k=0$ and is discussed immediately below, while S- turns to infinity.

{\em Solution S-1. ---}

If $k = 0$ in Eq.(3), we have a particular solution for EQ of an especially simple form:
\begin{equation}
A(r) = A_0\left(1 -  \frac{1}{\alpha r^2}\right), \;\; B(r) = r^2, \;\; 
C(r) =\frac{2}{\alpha r^2}\, \left(1 -\frac{1}{\alpha r^2}\right)^{-1};
\end{equation}
\begin{equation}
8\pi \rho =  \frac{3}{2}\, \alpha\, \left(\frac{1}{\alpha r^2} -1\right) \;,
\end{equation}
\noindent
where  $\alpha = - b $ and $A_0=a^2\alpha/2$. To keep the proper signature of the metric, we require $A_0>0$, so that $\alpha>0$; without loss of generality, we set $A_0=1$, i. e., $a^2=2/\alpha=-2/b>0$. Let us discuss this solution in more detail, setting, for simplicity, $\alpha=1$.

\noindent A) The space-time of the solution S-1 has a
horizon at $r = 1$ where the components of the metric
tensor $g_{00} = A(r)$ and $g_{11} = -C(r)$ change their signs.
The horizon separates (or connects)
a static spatially inhomogeneous exterior space (E-space)
at $r > 1$ and a `hidden' interior object (I-object) at $r < 1$. These two
regions are similar to R- and T-regions, respectively, as described by
Novikov [8] for the Schwarzschild solution.
The true singularity with the infinite Ricci curvature and density is at $r = 0$.
The density in S-1 may be regarded as consisting of two
components with the same equation of state: one is
uniform and negative, $8\pi\rho _{-}  = - 3/2 < 0$, the other is
nonuniform (`isothermal' law) and positive,
$8\pi\rho_+ (r) = 3/2r^{2} > 0$. The first one dominates
the E-space, and the other one dominates the I-object; the total
density changes its sign at the horizon.

\noindent B) The `energy equation' introduced for the metric of Eq.(1)
by Lema\^itre (see Ref. [5]) may be used to find an analog of the
Newtonian gravitational potential
as one of the physical characteristics of gravity produced by EQ in S-1.
The equation has a form of energy conservation relation,
$ \frac{1}{2} (dB^{1/2}/dt)^2 - m(r,t)/r = E$,
where $E = C^{-1} (dB^{1/2}/dr)^2 - 1$ and $m(r,t)$ is the Newtonian
analog of gravitating mass. In S-1, the Newtonian mass proves to
be $m(r) = (3/4) r [1 - (1/3) r^2]$, and the Newtonian potential
$u(r) = - m/r = - (3/4) [1 - (1/3) r^2]$ describes `the Newtonian
component' of gravity produced by EQ in S-1. The potential $u$ is finite at
the true
singularity [$u(0) = - 3/4$] (unlike that in Schwarzchild solution)
and goes to $+ \infty$ at spatial infinity where density becomes
constant (see below). On the singular surface,
$m(1) = 1/2$, and so $2|u(1)| = 1$ (like that in the Schwarzschild solution).

\noindent C) EQ gravity in S-1 is not reduced to the Newtonian
gravity completely; in particular, the dynamic effect of acceleration is
essentially different in S-1 from what may be produced by the Newtonian
potential $u$. The total accelerating effect can be evaluated
in terms of the effective potential, $U(r)$, which may be derived using
the fact that S-1 describes EQ in a state of hydrostatic equilibrium:
its pressure and self-gravity are balanced in the E-space.
The pressure gradient produces radial acceleration
$ F_p =  - (dp/dr)(\rho + p)^{-1} = [8\pi (r^3 - r)]^{-1}$,  which is
positive. So the gravity acceleration is
$F = - F_p = - [8\pi (r^3 - r)]^{-1} $; it is negative, and therefore
gravity is attracting. The effective gravitational potential responsible for
this attraction is $U =  (1/2) \ln \left[r^2/(r^2 - 1)\right]$.

\noindent D) S-1 can be rewritten in the E-space as
\begin{equation}
ds^2 = {\tanh ^2 (\chi /\sqrt{2})} dt^2 - \cosh ^2 (\chi/\sqrt{2})
d\Omega ^2 - d\chi ^2,
\end{equation}
\begin{equation}
8\pi \rho = - {\frac{3}{2}} \tanh (\chi/(\sqrt{2}),
\end{equation}
\noindent
\noindent
with the new spatial coordinate $\chi$ related to $r$ by
$r = \cosh (\chi/(\sqrt{2})$; its range is from zero to infinity.
The density decreases monotonically from zero at $\chi = 0$ to $- 3/2$
at $\chi =\infty$.  This form of S-1 may possibly be
considered both in relation to and independently of the I-object.
In the limit $r \rightarrow \infty$, where density is spatially
homogeneous, the metric of S-1 takes the form
$ds^2 =dt^2-r^2d\Omega^2-2r^{-2}dr^2$, or $ ds^2 =  dt^2 - \exp (\sqrt{2} \chi) d\Omega ^2 - d\chi ^2$.
Since the general static solution for EQ in the isotropic 3-space has the form
$ ds^2 =  dt^2 - r^2 d\Omega^2 -  (1 - k r^{2}/a_0 ^2)^{-1} dr^2$,
$ 8\pi \rho = - 24 \pi p = 3k/a_0 ^2$,
($k = 1, 0, - 1$ is the sign of 3-curvature, $a_0$ is
an arbitrary scalling constant), one can see that the metric of
Eqs.(2,4) reduces to the isotropic metric with $k = - 1$ and
$a_0 = 1/2$ in the limit $r  \rightarrow \infty$. For comparison:
$k = 1, a_0 = 1$ in the Einstein static solution.

\noindent E) Since $A(r)$ and $C(r)$ in S-1 change their
signs at the singular surface, the signature of the metric
of the I-object is (- - - +). This means that the coordinate $r$
becomes time-like, and the coordinate $t$ becomes space-like inside the
I-object. The same is true for the Schwarzschild space-time
([8]). Because of that, the density of the I-object
depends only on the time-like coordinate, thus it is spatially homogeneous.
However, in contrast with the
Schwarzschild space-time, a formal replacement of $g_{00}$ with $g_{11}$
and vice versa, which transforms the signature of the I-object to the
`ordinary' type (+ - - -), does not
work for S-1, because the metric resulting from Eq.(2) after this
transformation is not a solution to the GR equations.

{\em Solution S-2. }

GR equations for the same metric of Eq.(1) allow for
the following time-dependent particular solution:
\begin{equation}
A(t) = \frac{1}{4}\,(n^2-4) = {\rm const} > 0, \;\;\;
B(t) = t^2, \;\;\; C(t) = t^n,
\end{equation}
\begin{equation}
8\pi \rho =  \frac{n(n+4)}{n^2-4}\, \frac{1}{t^{2}}, 
\qquad p = w \rho,
\qquad n = - \frac{4 w}{1 + w}= {\rm const}.
\end{equation}
The parameter range of this solution apparently consists of two parts: $n>2$ ($-1<w<-1/3$), and $n<-2$ ($|w|>1$; this is not  Q as defined above). Note that both VQ ($w=-1$) and EQ ($w=-1/3$) are outside this range.
\noindent

\noindent A) The most striking feature of this solution (S-2) is that
$A, C$ and even $B$ do not depend on $r$, and are functions of the time only.
One may see here a similarity to the I-object
of S-1 whose metric also depends on the time-like coordinate only.
A more close similarity may be recognized with the
`T-sphere' found by Ruban [9] for pressure-free matter and the same as
in S-2 symmetry of 3-space with $B = B(t)$.
The density and pressure are spatially homogeneous (functions
of the time only) in both the S-2 and Ruban's T-sphere.
Time $t$ varies in S-2 from $- \infty $ to $+ \infty$,
and the density varies from zero at $t \rightarrow \pm \infty$ to infinity
in (true) singularity $t=0$ which has the same character as that in isotropic
cosmological models. However, there
are no coordinate transformations that could reduce S-2 and Ruban's
T-sphere to the FRW metric with the isotropic 3-space.

\noindent B) Unlike the Friedmann solutions, gravity in S-2 and
Ruban's T-sphere does not have any Newtonian analogs.
Rather, the space-time of this special type has common features
with anisotropic spatially homogeneous cosmological models (cf. [9]).

\noindent C) The pressure, $p=-n^2/[8\pi(n^2-4)t^2]$, is negative in S-2, while the density may be either positive or negative. The density is positive if $w$ is in `the Q range' $- 1 < w < - 1/3$ ($n>2$). (Note that in the isotropic 3-space Q with
$w$ in this range produces positive acceleration, see Introduction.)
The density is also positive for $w > 1$ (i. e., $-4<n< 2$), but it is negative when 
$w<-1$ ($n<-4$).

\noindent D) Density turns to zero for $w = \infty$ ($n=-4$). This case describes
a form of energy with $\rho = 0,\, p < 0$, which is perhaps the
most extreme form of Q. As we see, GR does not exclude
the equation of state with $w = \infty$, and gives a regular solution for it,
$A(r)  = 3, \;\; B(r) = t^2 \; \; C = t^{-4}$.
`Density-free' energy may also be in a state of hydrostatic
equilibrium (see below).

{\em Solution S-3. }

A power law particular static solution, which is a
counterpart to S-2, is also allowed by GR equations for the metric of Eq.(1):
\begin{equation}
A(r) = r^n, \;\;\; B(r) = r^2, \;\;\; C(r) =  \frac{1}{4}\,(4+4n-n^2)=
{\rm const} > 0,
\end{equation}
\begin{equation}
8\pi \rho = \frac{n(4 - n)}{4 + 4n - n^2}\,\frac{1}{r^{2}}, 
\qquad p = w \rho,
\qquad n = \frac{4 w}{1 + w}.
\end{equation}
\noindent
The parameter range here looks even more peculiar: $-2(\sqrt{2}-1)<n<2(\sqrt{2}+1)$, which corresponds to the two intervals for the values of $w$, 
$w>-(\sqrt{2}-1)/(\sqrt{2}+1)>-1/3$ and $w<-1$. Neither VQ, nor EQ are within this range, and the solution relates to Q only when $-(\sqrt{2}-1)/(\sqrt{2}+1)<w<0$.

\noindent A) Similar to S-1 and S-2, solution S-3 depends on one
coordinate only, and the true singularity of all the
three solutions is at the origin of this coordinate. The
positive component of density in S-1, and the total densities in S-2 and
S-3 follow the inverse square law.
Both S-1 and S-3  describe Q in a state of hydrostatic equilibrium.

\noindent B) The pressure, $p=n^2/[8\pi(4+4n-n^2)r^2]$, is positive in S-3 for any $n \not= 0$.
At $n = 0$ the pressure and density are both identical zeros. In this case,
$A = 1, C = 1$, and the metric
of S-3 turns to the Lorentz metric of the empty space-time.
The density in S-3 is positive for $0 < n < 4$ and negative
otherwise.

\noindent C) The density vanishes also for $ n = 4,\, w = \pm \infty $.
This is the `density-free' ($\rho = 0,\, p > 0$) form of energy, and S-3 (same as S-2 above)
gives a regular metric for this case: $A = 1, B(r) = r^2, C(r) = r^{4}$.
The fact that this form of energy is in the state of hydrostatic
equilibrium  shows that `density-free' energy has its
inertial mass (per unit volume) $\rho _i = p$ which is equal to its passive
gravitational mass, and that the
active mass is positive.

\noindent D) For $n \not= 0$, the sign of the active gravitating mass
depends on the sign of $n$. If $n$ is
positive, the mass is positive; but if $n$ is negative (and thus $w$ is negative),
the active mass is negative also.
In terms of Newtonian physics, the negative mass produces a positive
acceleration, which goes to infinity as
$r \rightarrow 0$, and so anti-gravity is infinitely strong at the
singularity. Whether it is capable, under these conditions, of producing
`auto-emission' of particles from the singularity (where the density is
infinite), and/or enhance quantum
evaporation of particles from the singularity, should be discussed separately.

\noindent E) S-3 describes hydrostatic equilibrium of not
only Q with negative $w$, but also, for instance, of ultra-relativistic fluid
($w = 1/3, n = 1, C = 7/4$,) and  of Zeldovich ultra-stiff fluid
($w = 1, n = 2, C = 2$).

\vspace{1,5cm}

[1] R.R. Caldwell, R. Dave, P.J. Steinhard, Phys.Rev.Lett. 80, 1582 (1998)

[2] R.R. Caldwell, R. Dave, P.J. Steinhard, Ap. Space Sci. 261, 303 (1998);
L. Wang, P.J. Steinhard, Astrophys. J. 508, 483 (1998);
C.-P. Ma, R.R. Caldwell, P. Bode, L. Wang, Astrophys. J. 521, L1 (1999);
A.R. Cooray, D. Huterer, Astrophys. J. 513, L95 (1999);
J.S. Alcaniz, J.A.S. Lima, Astron. Astrophys. 349, 729 (1999);
I.S. Zlatev, P.J. Steinhard, Phys. Lett. B 459, 570 (1999);
L. Hui, Astrophys. J. 519, L9 (1999);
I. Zlatev, Wang L., P.J. Steinhard, Phys. Rev. Lett. 82, 896 (1999);
P.J.E. Peebles, A. Vilenkin, Phys. Rev. D 590, 811 (1999);
M. Giovannini, Phys. Rev. D 601, 277 (1999);
L. Wang, R.R. Caldwell, J.P. Ostriker, P.J. Steinhard, Astrophys. J.
550, 17 (200); G. Efstathiou, MN RAS 310, 842 (2000);
P.F. Gonz\'alez-D\'iaz, Phys. Lett. B 481, 353 (2000);
J.D. Barrow, R. Bean, J. Magueijo, MN RAS 316, L41 (2000)

[3] S.M. Carroll, Phys. Rev. Lett. 81, 3067 (1998);
S.M. Barr, Phys. Lett. B 454, 92 (1999);
R.S. Kalyana, Phys. Lett. B 457, 268 (1999);
Ch. Kolda, D.H. Lyth, Phys. Lett. B 459, 570 (1999);
P. Bin\'etruy, Phys. Rev. D 600, 80 (1999);
R. Horvat, Mod. Phys. Lett. A 14, 2245 (1999);
T. Chiba, Phys. Rev. D. 601, 4634 (1999);
P.H. Brax, J. Martin, Phys. Lett. B 468, 40 (1999);
A.B. Kaganovich, Nuc. Phys. B Proc.Suppl. 87, 496 (1999);
O. Bertolami, P.J. Martins, Phys. Rev. D 610, 7 (2000);
Y. Nomura, T. Watari, T. Yanagida, Phys. Lett. B 484, 103 (2000);
S.C.C. Ng, Phys. Lett. B 485, 1 (2000); Dynmikova I. G. Phys. Lett. B472, 33 (2000);
A. Hebecker, C. Wetterich, Phys. Rev. Lett. 85, 3339 (2000);
N. Arkani-Hamed, L.J. Hall, C. Colda, H. Murayama, Phys. Rev. Lett. 85,
4434 (2000);
C. Armendariz-Picon, V. Mukhanov, P.J. Steinhard, Phys. Rev. Lett. 85,
4438 (2000)

[4] S. Weinberg, Rev. Mod. Phys. 61, 1 (1989)

[5] G. Lema\^itre, Rev. Mod. Phys. 21, 357 (1949);
E.B. Gliner, JETP 22, 378 (1966); E.B. Gliner, Sov. Phys. Doklady
6, 559 (1970).

[6] A.G. Riess et al., Astron. J. 116, 1009 (1998);
S. Perlmuter et al., Astrophys. J. 517, 565 (1999);
J. Cohn, astro-ph/9807128;
S. Carol, astro-ph/0004075

[7] L.D. Landau, E.M. Lifshitz, The Classical Field Theory, Pergamon Press,
London-Paris (1959)

[8] I.D. Novikov, Comm. Sternberg Astron. Inst. 132, 43 (1963)

[9] V.A. Ruban, JETP 29, 1027 (1969)

\end{document}